\documentclass[12pt]{article}
\usepackage{amsmath}
\usepackage[dvips]{graphicx}

\begin{document}
\begin{center}
\textbf{\large Running of the Spectral Index and Violation of the
Consistency Relation Between Tensor and Scalar Spectra from
trans-Planckian
Physics}\\[0pt]

\vspace{48pt} A. Ashoorioon\footnote{%
amjad@astro.uwaterloo.ca}, J. L. Hovdebo\footnote{%
jlhovdeb@sciborg.uwaterloo.ca}, R. B. Mann\footnote{%
mann@avatar.uwaterloo.ca}

\vspace{12pt}

\textit{Department of Physics, University of Waterloo, \\[0pt]
Waterloo, Ontario, N2L 3G1, Canada }

\vspace{24pt}

\begin{abstract}
\addtolength{\baselineskip}{1.2mm} \addtolength{\baselineskip}{1.2mm} One of
the firm predictions of inflationary cosmology is the consistency relation
between scalar and tensor spectra. It has been argued that such a relation
--if experimentally confirmed -- would offer strong support for the idea of
inflation. We examine the possibility that trans-Planckian physics violates
the consistency relation in the framework of inflation with a cut-off
proposed in astro-ph/0009209. We find that despite the ambiguity that exists
in choosing the action, Planck scale physics modifies the consistency
relation considerably. It also leads to the running of the spectral index.
For modes that are larger than our current horizon, the tensor spectral
index is positive. For a window of $k$ values with amplitudes of the same
order of the modes which are the precursor to structure formation, the
behavior of tensor spectral index is oscillatory about the standard Quantum
field theory result, taking both positive and negative values. There is a
hope that in the light of future experiments, one can verify this scenario
of short distance physics.
\end{abstract}
\end{center}

\addtolength{\baselineskip}{1.5mm} \thispagestyle{empty}

\vspace{48pt}

\setcounter{footnote}{0}

\newpage

\section{Introduction}

One of the intriguing properties of inflationary cosmology, which could be
used to test the fundamental theories of quantum gravity, is its capacity to
accommodate sub-Planckian fluctuations that were redshifted exponentially
during a quasi-de-Sitter expansion of the universe \cite{Brandenberger1}.
Many realizations of inflation predict several more e-foldings than are
required to solve the problems of standard cosmology \cite{Linde}. Assuming
that these inflationary models are correct, all scales of cosmological
interest today originate inside the Planck scale at the early stages of
inflation. These fluctuations would be manifest in the temperature
anisotropy of the cosmic microwave background radiation (CMBR), which can be
regarded as a fossil record of primeval inhomogeneities. It is therefore
reasonable to expect that by studying the cosmic microwave background
radiation one can extract information about physics at very small distance
scales \cite{Greene}.

Several approaches have been employed to find out how modifications
of Planck scale physics affect orthodox calculations of the power
spectrum in the context of Quantum Field Theory. Originally,
Brandenberger and Martin considered the effect of Planck scale
physics to be encoded in modifications to dispersion relations
\cite{Brandenberger2}. Similar methods were employed to investigate
the same phenomenon in the context of black hole physics
\cite{Unruh,Corely}. These modifications were inspired by higher
dimensional models of the universe \cite{Riotto} or from condensed
matter analogs of gravity \cite{Volovik}. It was shown that the
prediction of a thermal Hawking spectrum is insensitive to
modifications of the physics at the trans-Planckian end of the
spectrum.

In an inflationary setting Planck scale physics may or may not leave an
imprint, depending on whether the mode behaves adiabatically when its
wavelength is smaller than the cut-off scale. Nonadaibatic evolution of a
mode when its wavelength is smaller than the Planck scale results in an
excited state at the time that the wavelength crosses the Hubble radius
during inflation \cite{Brandenberger3}. However, since this leads to a large
amount of particle production by trans-Planckian physics, it has been
claimed that such a possibility is excluded \cite{Starobinsky,Tanaka}.

Danielsson proposed a method in which our lack of knowledge about Planck
scale physics is parametrized in the choice of state at the time a mode
reaches the minimum scale \cite{Danielsson}. The state at the minimum
scale-crossing is the state of minimized uncertainty. Modifications to the
standard analysis are of order $H/\Lambda $, where $H$ is the Hubble
parameter during inflation and $\Lambda $ is the scale at which new physics
appears. This scenario was later generalized to power-law backgrounds in \cite%
{alberghi}.

A beautiful mechanism was suggested in \cite{Kempf1} to incorporate minimum
length into the inflationary formalism. The only assumption underlying this
formalism is that the fundamental theory of quantum gravity possesses a
linear operator $X^{i}$ for every space-time coordinate and that its
expectation value $\langle X^{i}\rangle $ is real. One can then show that
the short distance structure of any such coordinate could not only be
continuous or discrete, but could also be unsharp in one of two ways \cite%
{Kempf2}. The two unsharp cases are distinguished by the so-called
deficiency indices of $X^{i}$ being either nonzero and equal (Fuzzy type A)
or unequal (Fuzzy type B) \cite{Kempf2}. In the case of fuzzy type B,
sequences of vectors in the physical domain exist such that $\triangle X^{i}$
converges to zero. They are fuzzy in the sense that vectors of increasing
localization around different expectation values in general do not become
orthogonal. Fuzzy type A behavior has appeared in a number of studies in
quantum gravity and string theory where the uncertainty in $\triangle X^{i}$
has a finite lower bound at Planck scale \cite{Gross}. This short distance
structure can be modelled as quantum gravitational correction to the
commutation relation between the position and momentum operators
\begin{equation}
\lbrack \mathbf{X},\mathbf{P}]=i(1+\beta \mathbf{P}^{2}),
\label{commutation}
\end{equation}%
where $\beta ^{1/2}$ parameterizes the minimum length. The equations for
tensor modes were later analyzed numerically in \cite{Easther2,Easther3},
and it was predicted that the effect on the CMBR can be as large as $\sigma $%
, where $\sigma $ is the ratio of the minimum length to the Hubble length
during inflation, $\beta ^{1/2}H\equiv \sigma $.

Quite recently it was discovered in $\cite{Ambiguity}$ that this mechanism
of implementing minimal length in the action has an ambiguity: The usual
strategy for determining the initial condition requires reformulating the
action and discarding a boundary term. In the absence of minimal length, two
actions that differ by a boundary term are equivalent. However, the
introduction of a minimal length scale renders two actions that normally
differ by a boundary term inequivalent, yielding different equations of
motion. One has an infinite set of actions that are equivalent when the
minimal length is set to zero. Only experiment can adjudicate which choice
of action is preferable. Nevertheless, in \cite{tensor/scalar}, from the
infinite number of actions that are equivalent in the absence of minimal
length, we adopt two actions for each of tensor and scalar fluctuations. The
first one is chosen by a minimalist criterion: we select the action that is
derived directly by expanding the action of a scalar field minimally coupled
to gravity without introducing any additional terms by hand. The second
action, which differs from the first by a boundary term, is chosen by the
criterion of similarity with the action of a free massive scalar field in a
Minkowskian background. Such a similarity simplifies the task of choosing
the vacuum. Basically, one can choose the vacuum as one does in Minkowskian
space-time.

A simple, near-de-Sitter background has recently been investigated in this
context \cite{tensor/scalar}, where it was shown that the tensor/scalar
ratio, gets modified as one incorporates minimal length. Trans-Planckian
physics may or may not leave its thumbprint on this ratio depending on the
actions one chooses for tensor and scalar perturbations. Such an ambiguity
was also employed in \cite{cosmag} to account for the existence of cosmic
magnetic fields. In this article, we consider the implications of minimal
length in a power-law background to find the possible scale-dependence of
the tensor/scalar ratio.

We will examine, in the context of the minimal length hypothesis as
implemented in \cite{Kempf1}, a firm prediction of inflationary cosmology,
the consistency relation between scalar and tensor perturbations. In the
case of single-field slow-roll inflation the consistency conditions are
given in terms of equality relations, whereas for multiple-field models of
inflation these are weakened to inequalities. The first of the consistency
relations states that the ratio of the amplitude of tensor to scalar
perturbations is a constant known as the tensor spectral index \cite%
{star-lyth}. We investigate how the effects of trans-Planckian physics alter
this ratio and the tensor spectral index under the considerations noted in %
\cite{Kempf1}. This is in contrast to recent work in this area in which this
possibility was investigated without focusing on any specific model of short
distance physics, instead assuming that the trans-Planckian energies result
in a vacuum state that is different from the standard Bunch-Davies vacuum %
\cite{Kinney}. We shall restrict ourselves to single-field inflation for the
rest of the discussion, although our results could straightforwardly be
generalized to multiple-field inflation.

Our paper is structured as follows: first we recapitulate our results \cite%
{Ambiguity} for both tensor and scalar fluctuations. Next, we study
numerically the equations of motion for scalar and tensor modes in a
power-law background and derive the tensor/scalar fluctuations and the
tensor spectral index in each case. As mentioned earlier, actions that
differ by a boundary term are rendered inequivalent once one implements the
minimal length hypothesis. Although this implies an infinite amount of
freedom in choosing the action for both scalar and tensor fluctuations,
there are only a few actions that have reasonable physical motivation, and
we shall confine our considerations to these cases. Specifically, we shall
discuss how these physically well-motivated but distinct actions modify the
consistency relation between tensor and scalar spectra.

\section{Scalar and Tensor Perturbations with Minimum Length}

To find the action for scalar and tensor perturbations, we expand
the action of a scalar field minimally coupled to gravity
\begin{equation}
S=\frac{1}{2}\int (\partial _{\mu }\phi \partial ^{\mu }\phi -V(\phi ))\sqrt{%
-g}d^{4}x-\frac{1}{16\pi G}\int R\sqrt{-g}d^{4}x  \label{2}
\end{equation}%
using the most general form of the metric with scalar and tensor
fluctuations
\begin{equation}
ds^{2}=a^{2}(\tau )\left[ (1+2\Phi ){d\tau }^{2}-\right( (1-2\Psi ){\delta }%
_{ij}+h_{ij}\left) d{y}^{i}d{y}^{j}\right] .  \label{mtrc}
\end{equation}%
where $\Phi $ and $\Psi $ are scalar fields and $h_{ij}$ is a symmetric
tensor field satisfying transverse traceless gauge, $%
h_{i}^{i}=h_{ij}{}^{,j}=0$. At the same time, we also perturb the inflaton
field about its homogeneous background value
\begin{equation}
\phi (\tau )=\phi _{0}(\tau )+\delta \phi .  \label{inflat}
\end{equation}%
Here $\phi _{0}$ is the homogeneous part that drives inflation and $\delta
\phi \ll \phi _{0}$. Also $\tau $ is the conformal time and $a\left( \tau
\right) $ the scale factor of the inflating spatially flat background.

Using Eqs.(\ref{mtrc}) and (\ref{inflat}) to expand action
(\ref{2}) to second order, the action for scalar perturbations can
be written in terms of the intrinsic curvature perturbations of
the comoving hypersurface, $\Re =-\frac{a^{\prime
}}{a}\frac{\delta \phi }{\phi _{0}^{\prime }}-\Psi $, in the
following form \cite{Ambiguity}
\begin{equation}
S_{S}^{(1)}=\frac{1}{2}\int d\tau ~d^{3}\mathbf{y}~z^{2}\left( (\partial
_{\tau }\Re )^{2}-\delta ^{ij}~\partial _{i}\Re \partial _{j}\Re \right) ,
\label{scr1}
\end{equation}%
where $\partial _{i}$ denotes differentiation with respect to spatial
coordinates and
\begin{equation}
z=\frac{a\phi _{0}^{\prime }}{\alpha },~~~~\alpha =a^{\prime }/a.  \label{z}
\end{equation}%
The prime denotes differentiation with respect to $\tau $. The
quantity $\Re $ is a gauge-invariant combination of scalar
fluctuations of the metric and inflaton perturbations
\cite{Lidsey}.

Incorporating the minimal length hypothesis involves retaining the
action (\ref{2}) as above, but modifying the underlying
position-momentum commutation relation similar to
eq.(\ref{commutation}). Specifically, the first order form of the
commutation relation in $\beta $ has the form \cite{Kempf1}:
\begin{equation}
\lbrack \mathbf{X}^{i},\mathbf{P}^{j}]=i\left( \frac{2\beta p^{2}}{\sqrt{%
1+4\beta p^{2}}-1}\delta ^{ij}+2\beta \mathbf{P}^{i}\mathbf{P}^{j}\right)
\label{27}
\end{equation}%
whose Hilbert space representation can be conveniently written as
\begin{equation}
\mathbf{X}^{i}\psi (\rho )=i\partial _{\rho ^{i}}\psi (\rho )~~~~~~~~\mathbf{P}^{i}\psi (\rho )=\frac{\rho ^{i}}{1-\beta \rho ^{2}}%
\psi (\rho )
\end{equation}%
where the scalar product of two quantum fields is
\begin{equation}
\left( \psi _{1},\psi _{2}\right) =\int_{\rho ^{2}<\beta
^{-1}}d^{3}\rho ~\psi _{1}^{\ast }(\rho )\psi _{2}(\rho )
\label{29}
\end{equation}%
It is then straightforward to derive the cutoff modified equation of motion
for the fluctuation mode $u_{\tilde{k}}$, which is \cite{Ambiguity}%
\begin{equation}
{u}_{\tilde{k}}^{\prime \prime }+\frac{{\kappa }^{\prime }}{{\kappa }}{u}_{%
\tilde{k}}^{\prime }+\left( \mu -\frac{z^{\prime \prime }}{z}-\frac{%
z^{\prime }\kappa ^{\prime }}{z\kappa }\right) {u}_{\tilde{k}}=0,
\label{scr1-eq}
\end{equation}%
where $u_{\tilde{k}}\equiv z\Re _{\tilde{k}}$ and
\begin{eqnarray}
\mu (\tau ,\tilde{k}) &=&-\frac{a^{2}}{\beta }\frac{W({-\beta \tilde{k}%
^{2}/a^{2}})}{(1+W({-\beta \tilde{k}^{2}/a^{2}}))^{2}}  \label{36} \\
\kappa (\tau ,\tilde{k}) &=&\frac{e^{-\frac{3}{2}W({-\beta \tilde{k}%
^{2}/a^{2}})}}{1+W({-\beta \tilde{k}^{2}/a^{2}})}.
\end{eqnarray}%
Here, $W$ is the Lambert $W$-function, defined via $W(x)e^{W(x)}=x$ \cite%
{Corless} and $\tilde{k}^{i}=a\rho ^{i}e^{-\beta \rho ^{2}/2}$ where $\rho
^{i}$ is the Fourier transform of the physical coordinate $x^{i}$. $\tilde{k}%
^{i}$ is a variable that is equivalent to comoving momentum at large
wavelengths.

Historically, the gauge invariant parameter $u$ was introduced to rewrite $%
S_{S}^{(1)}$ in the form of an action for a scalar field with a
time-dependent mass $z^{\prime \prime }/z$ in Minkowskian space-time \cite%
{Mukhanov}:
\begin{equation}
S_{S}^{(2)}=\frac{1}{2}\int d\tau d^{3}\mathbf{y}\left( {\left( \partial
_{\tau }u\right) }^{2}-\delta ^{ij}~{\partial }_{i}u~{\partial }_{j}u+\frac{%
z^{\prime \prime }}{z}u^{2}\right)   \label{scr2}
\end{equation}%
The resemblance of $S_{S}^{(2)}$ with the action of a massive scalar field
in a Minkowskian background simplifies the task of choosing the initial
conditions. Hence, the vacuum can be chosen following a similar procedure to
that in Minkowskian space-time. The equation of motion derived from the
cutoff modified $S_{S}^{(2)}$ is:
\begin{equation}
{u}_{\tilde{k}}^{\prime \prime }+\frac{{\kappa }^{\prime }}{{\kappa }}{u}_{%
\tilde{k}}^{\prime }+\left( \mu -\frac{z^{\prime \prime }}{z}\right) {u}_{%
\tilde{k}}=0.  \label{scr2-eq}
\end{equation}%
The difference between $S_{S}^{(1)}$ and $S_{S}^{(2)}$ is a boundary term
\begin{equation}
\triangle S_{S}\equiv S_{S}^{(1)}-S_{S}^{(2)}=\int d\tau ~d^{3}\mathbf{y}~%
\frac{d}{d\tau }\left( \frac{z^{\prime }}{z}u^{2}\right)   \label{diff-scr}
\end{equation}%
that can be discarded as long as we have not implemented the cutoff.

However the minimal length hypothesis transforms the boundary term (\ref%
{diff-scr}) in the following manner in $\tilde{k}$ space \cite{Ambiguity}%
\begin{equation}
\triangle S_{S}\rightarrow \int d\tau d^{3}\tilde{k}~\kappa (\tau ,\tilde{k}%
)~\frac{d}{d\tau }\left( \frac{z^{\prime }}{z}u^{2}\right) ,
\label{var-bndry}
\end{equation}%
or equivalently it adds
\begin{equation}
-\int d\tau d^{3}\tilde{k}\frac{d\kappa (\tau ,\tilde{k})}{d\tau }~\left(
\frac{z^{\prime }}{z}u^{2}\right)   \label{var-blk}
\end{equation}%
to the bulk term of the Lagrangian. The difference between eqs. (\ref%
{scr1-eq}) and (\ref{scr2-eq}) arises from the variation of the additional
term (\ref{var-blk}).

Following \cite{Lidsey}, we define the scalar amplitude as
\begin{equation}
A_{S}(k)\equiv \frac{2}{5}P_{S}^{1/2}=\frac{2}{5}\sqrt{\frac{k^{3}}{2\pi ^{2}%
}}\left| \frac{u_{\tilde{k}}}{z}\right| _{\tilde{k}/aH\rightarrow 0}.
\label{scrpower}
\end{equation}%
Perturbing the Einstein-Hilbert action about a homogenous,
isotropic, spatially flat background as in (\ref{mtrc}), one can
write down the action for tensor perturbations in the transverse
traceless gauge as \cite{Lidsey}
\begin{equation}
S_{T}^{(1)}=\frac{m_{Pl}^{2}}{64\pi }\int d\tau d^{3}\mathbf{y}~a^{2}(\tau
)~\partial _{\mu }h^{i}{}_{j}~\partial ^{\mu }h_{i}{}^{j}.  \label{tsr1}
\end{equation}%
As in the scalar case, one can rewrite $S_{T}^{(1)}$ in the form of an
action for a tensor field with time dependent mass in a Minkowski background
by introducing the new variable $P^{i}{}_{j}(y)\equiv \sqrt{\frac{m_{Pl}^{2}%
}{32\pi }}~a(\tau )h^{i}{}_{j}(y)$
\begin{equation}
S_{T}^{(2)}=\frac{1}{2}\int d\tau d^{3}\mathbf{y}\left( \partial _{\tau
}P_{i}{}^{j}\partial ^{\tau }P^{i}{}_{j}-\delta ^{rs}{\partial }%
_{r}P_{i}{}^{j}{\partial }_{s}P^{i}{}_{j}+\frac{a^{\prime \prime }}{a}%
P_{i}{}^{j}P^{i}{}_{j}\right) .  \label{tsr2}
\end{equation}%
where $S_{T}^{(2)}$ differs from $S_{T}^{(1)}$ by a boundary term%
\begin{equation}
\triangle S_{T}\equiv S_{T}^{(2)}-S_{T}^{(1)}=\frac{32\pi }{m_{Pl}^{2}}\int
d\tau d^{3}\mathbf{y}\left( \alpha P_{i}{}^{j}~P^{i}{}_{j}\right) ^{\prime }.
\label{tsr-diff}
\end{equation}

Incorporating the minimal length hypothesis (\ref{commutation}) into $%
S_{T}^{(1)}$ and $S_{T}^{(2)}$ respectively yields the following equations
of motion for the $\tilde{k}$-Fourier transform of $P_{ij}$ $\ \cite%
{Ambiguity}$
\begin{equation}
{p}_{\tilde{k}}^{\prime \prime }+\frac{{\kappa }^{\prime }}{{\kappa }}{p}_{%
\tilde{k}}^{\prime }+\left( \mu -\frac{a^{\prime \prime }}{a}-\frac{%
a^{\prime }}{a}\frac{{\kappa }^{\prime }}{{\kappa }}\right) {p}_{\tilde{k}%
}=0,  \label{tsr1-eq}
\end{equation}%
\begin{equation}
{p}_{\tilde{k}}^{\prime \prime }+\frac{{\kappa }^{\prime }}{{\kappa }}{p}_{%
\tilde{k}}^{\prime }+\left( \mu -\frac{a^{\prime \prime }}{a}\right) {p}_{%
\tilde{k}}=0.  \label{tsr2-eq}
\end{equation}%
These equations of motion differ because the minimal length hypothesis
implies
\begin{equation}
\triangle S_{T}\rightarrow \int d\tau d^{3}\tilde{k}~\kappa (\tau ,\tilde{k}%
)~\frac{d}{d\tau }\left( \alpha P_{i}{}^{j}~P^{i}{}_{j}\right)
\label{var-bndr-tsr}
\end{equation}%
thereby modifying the boundary term eq.(\ref{tsr-diff}) in a nontrivial
manner \cite{Ambiguity}. Following \cite{Lidsey}, we define the tensor
amplitude as
\begin{equation}
A_{T}(k)\equiv \frac{1}{10}P_{T}^{1/2}=\frac{1}{10}\sqrt{\frac{k^{3}}{2\pi
^{2}}}\left| h_{\tilde{k}}\right| _{\tilde{k}/aH\rightarrow 0} \ .
\label{tsrpower}
\end{equation}

\section{Tensor/Scalar Ratio and the Violation of the Consistency Relation}

One can expand the ratio of tensor/scalar fluctuations in terms of the slow
roll parameters in the absence of a cut-off. To first order it is \cite%
{star-lyth,Lidsey}
\begin{equation}
r\equiv \frac{A_{T}^{2}}{A_{S}^{2}}=\epsilon  \label{r-slowroll}
\end{equation}%
where
\begin{equation}
\epsilon \equiv \frac{3\dot{\phi}_{0}^{2}}{2}{\left( V(\phi _{0})+\frac{1}{2}%
{\dot{\phi}_{0}}^{2}\right) }^{-1}=\frac{m_{Pl}^{2}}{4\pi }\left( \frac{%
H_{\phi }}{H}\right) ^{2}  \label{eps}
\end{equation}%
with the $\phi $ subscript and over-dot respectively denoting
differentiation with respect to $\phi $ and the cosmic time, $t$, related to
conformal time $\tau $ by $t=\int ad\tau $. Both tensor and scalar
fluctuations contribute to the anisotropy of the CMBR. Hence, to extract the
characteristics such as spectral indices for each type of fluctuation we
need to know $r$ \cite{Salopek,Liddle}.

Since scalar and tensor perturbations originate from a single inflaton
potential they are not independent. A hierarchy of consistency conditions
links them together \cite{Lidsey}. It has been argued that such conditions
-- if empirically verified -- would offer strong support for the idea of
inflation. Observational difficulties will probably render only the first
consistency condition useful. The first of these consistency relations
relates $r$ to the tensor spectral index, $n_{T}$, defined as

\begin{equation}
n_{T}(k)\equiv \frac{d\ln A_{T}^{2}(k)}{d\ln k}.  \label{tsr-indx}
\end{equation}%
To first order in slow-roll parameters $n_{T}$ can expanded, yielding%
\begin{equation}
n_{T}=-2\epsilon,  \label{nt-eps}
\end{equation}%
and so the first-order consistency relation takes the following form
\begin{equation}
r\equiv \frac{A_{T}^{2}}{A_{S}^{2}}=-\frac{n_{T}}{2}  \label{consis}
\end{equation}%
in the absence of a cut-off.

In presence of minimal length the relation (\ref{r-slowroll}) is modified
\begin{equation}
\frac{A_{T}^{2}}{A_{S}^{2}}=\epsilon {\left| \frac{p_{k}}{u_{k}}\right| }%
_{k/aH\rightarrow 0}^{2}  \label{r-slowroll-gen}
\end{equation}%
where $u_{k}$ and $p_{k}$ will satisfy different differential equations
contingent upon the choice of action in the presence of a cutoff.
Furthermore, eq.(\ref{nt-eps}) no longer holds true\footnote{%
We are grateful to A. Kempf for bringing this to our attention}.

Hence one expects that Planck scale physics will modify the consistency
relation. Our predictions of course depend on the choice of the action for
tensor and scalar perturbations. As noted above, for both tensor and scalar
spectra there are two physically motivated actions, yielding four cases of
interest that we will separately analyze below.

\subsection{The Mode Equations in a Power Law Background}

Before presenting our numerical results for the power spectra, it will be
instructive to consider the explicit form of the mode equations (\ref%
{scr1-eq},\ref{scr2-eq},\ref{tsr1-eq},\ref{tsr2-eq}). A power-law
inflationary background is described by
\begin{equation}
a(t)=t^{p},~~~~~~~a(\tau )={\left( \frac{\tau }{\tau _{0}}\right) }%
^{q},~~~~~~~q=\frac{p}{1-p},  \label{scale}
\end{equation}%
where $t(\tau )$ is the cosmic (conformal) time and $p>1$. Assuming that at $%
t=1$, $a(t)=1$ then $\tau _{0}=1/(p-1)$. We will track the evolution of the
modes numerically from when they are created at the time $\tau _{\tilde{k}%
}\equiv \tau _{0}\left( e\beta \tilde{k}^{2}\right) ^{1/2q}$, until $\tau
\rightarrow 0$ at which point we calculate the power spectrum. To this end,
we define a new variable, $y$, so that $\tau =\tau _{\tilde{k}}(1-y)$. \ It
will prove convenient to work with the rescaled quantity $k=\tilde{k}e^{p/2}/%
\tilde{k}_{crit}$., where $\tilde{k}_{crit}$ corresponds to the mode that
crosses the horizon just before the Hubble radius reaches the minimal length
scale, $\sqrt{\beta }H=1$. Explicitly it is given by%
\begin{equation}
\tilde{k}_{crit}=e^{-1/2}p{(\beta p^{2})}^{(p-1)/2}.  \label{kcrit}
\end{equation}

With these definitions, the mode equation (\ref{scr1-eq}) becomes%
\begin{eqnarray}
\ddot{u}_{\tilde{k}}-\frac{q}{1-y}\frac{W(5+3W)}{(1+W)^{2}}\dot{u}_{\tilde{k}%
}-\left( \frac{eq^{2}k^{2/p}W}{(1-y)^{-2q}(1+W)^{2}} \right .  \notag \\
\left . +\frac{q(q-1)}{(1-y)^{2}}+\frac{q^{2}}{(1-y)^{2}}\frac{W(5+3W)}{%
(1+W)^{2}}\right) u_{\tilde{k}} & = & 0,  \label{mode_eq_explicit}
\end{eqnarray}%
%
where an overdot now denotes a derivative with respect to $y$, and the
argument of the Lambert W function is $-e^{-1}(1-y)^{-2q}$. The other mode
equations (\ref{scr2-eq}),(\ref{tsr1-eq}) and (\ref{tsr2-eq}) may be
obtained by dropping the final term in the parentheses and replacing $u_{%
\tilde{k}}$ with $p_{\tilde{k}}$ as necessary. The definition of $k$ was
chosen to remove the explicit dependence upon the minimal length, but we now
see that there is an added benefit to the choice of these variables. For
large $p$, to very good approximation $q$ is $-1$ . In fact, actually
setting $q=-1$ changes the equations very little. To a very good
approximation then, all the important dependence upon $p$ occurs through the
factor $k^{2/p}.$ When written using the $k$ variable the behavior of the
modes is independent of $\beta$, but following all the factors through we
find that the normalization of the power spectrum varies as $\beta^{-1/2}$.

We start tracking the mode numerically just after it is created by solving
the mode equations approximately for small $y$. The approximate solution
enables us to choose our vacuum, and fix the normalization of the mode
function. There is a branch cut in the Lambert W function when its argument
is $-e^{-1}$, but since we will start following the mode numerically from
some small but positive $y$ we may treat it as a removable singularity when
we determine the approximate solution, since $W\sim -1+2\sqrt{-qy}+O(y)$ as $%
y\rightarrow 0^{+}.$ In general the asymptotic solution is expressible in
terms of Hankel functions. The explicit form is dependent on the exact mode
equation since it is the terms having the factor $(1+W)^{-2}$ that dominate
as $y\rightarrow 0$.

\subsection{\protect\bigskip $(S_{T}^{(1)},S_{S}^{(2)})$}

Let us first assume that the actions for tensor and scalar fluctuations are $%
S_{T}^{(1)}$ and $S_{S}^{(2)}$, respectively. Since in this case $u_{\tilde{k%
}}$ and $p_{\tilde{k}}$ satisfy different equations, the tensor/scalar ratio
differs from the standard quantum field theory prediction. Solving equation (%
\ref{mode_eq_explicit}) near $y=0$ with the method of dominant balance \cite%
{Bender} (previously employed in other studies \cite%
{Easther2,Easther3,tensor/scalar}); we find
\begin{equation}
p_{k}(y)=D_{+}~G(k,y)(1+\xi _{1}(k,y))(1+\xi _{2}(k,y))+D_{-}~G^{\ast
}(k,y)(1+\xi _{1}^{\ast }(k,y))(1+\xi _{2}^{\ast }(k,y))  \label{sol-S1}
\end{equation}%
where%
\begin{eqnarray}
G(k,y) &=&y^{3/4}H_{-3/4}(2\sqrt{A_{k}y})  \notag  \label{sol1} \\
\xi _{1}(k,y) &=&-\frac{ek^{2/p}}{6}(-qy)^{3/2}(2-3\log y) \\
\xi _{2}(k,y) &=&\frac{qy^{2}}{48}\left(
3q(16-59ek^{2/p})+4i(2+ek^{2/p})(3i+7q^{2}\sqrt{2+ek^{2/p}})+42qek^{2/p}\log
y\right)  \notag
\end{eqnarray}%
and
\begin{equation}
A_{k}=-\frac{q}{4}\left( 2+ek^{2/p}\right) .  \label{ak}
\end{equation}

The quantities $D_{-}$ and $D_{+}$ are constrained by the Wronskian
condition which implies:
\begin{equation}
{\left\vert D_{+}\right\vert }^{2}-{\left\vert D_{-}\right\vert }^{2}=-\eta
_{\tilde{k}}\pi \sqrt{-q}{e}^{-3/2}.  \label{wronskian1}
\end{equation}%
In a near de-Sitter background if $D_{-}/D_{+}$ goes to zero when $\beta
\rightarrow 0$, then the standard QFT result can be recovered \cite%
{tensor/scalar}. However if $D_{-}/D_{+}$ is constant in this limit, one
cannot recover the standard QFT result as $\beta \rightarrow 0$ \cite%
{Easther3}. We conjecture that the same type of reasoning is valid in a
power-law background, and therefore we still have freedom in choosing the
vacuum. We shall proceed with the choice $D_{-}=0$, which corresponds to a
Bunch-Davies-like vacuum.

\begin{figure}[t]
\includegraphics[angle=0,
scale=0.80]{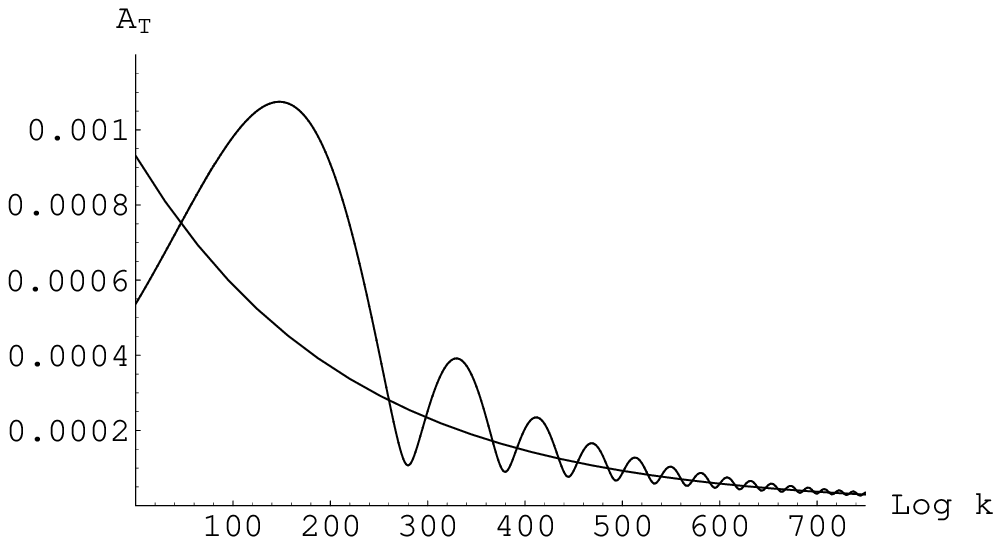}
\includegraphics[angle=0,
scale=0.80]{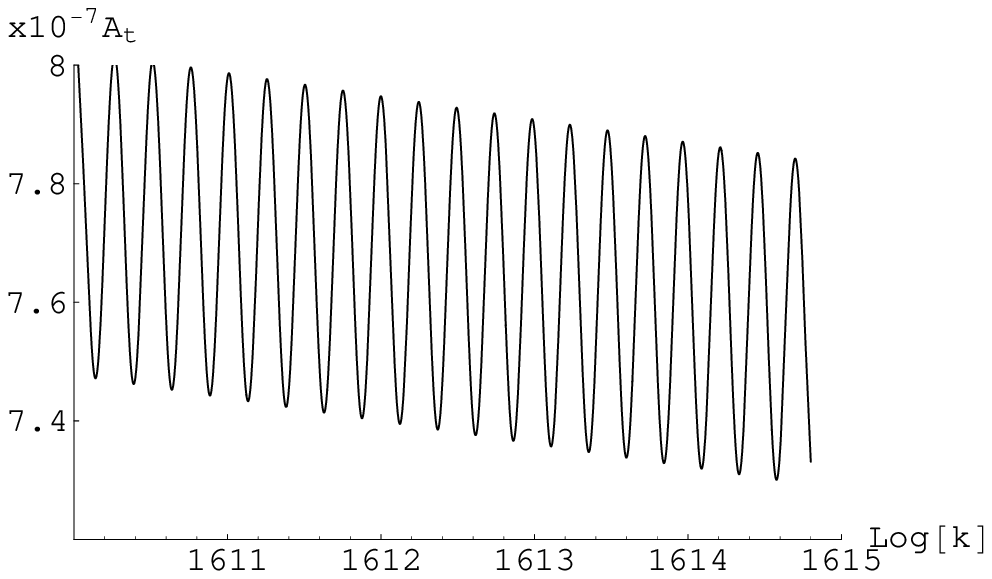}
\caption{In these figures we assume that tensor perturbations are described
by $S_{T}^{(1)}$. The left graph shows the dependence of $A_{T}$ on $\log k$
for $p=500$ and $\protect\beta =10^{4}$. $k=1$ corresponds to $k_{crit}(500)$%
. The large modulation corresponds to physical scales much larger
than our horizon. On the right, we plot the tensor amplitude with a
window of $k$ whose amplitudes are of the same order as the modes
that originated the structure formation in our universe.}
\label{fig1}
\end{figure}

This analytic solution can then be used as an initial condition to
numerically integrate the differential equation from a point in the vicinity
of the singular point until $\tau \approx 0.$ At this point, we extract the
late time amplitude of $u_{\tilde{k}}$. Figure \ref{fig1} illustrates our
results for the tensor amplitude for $p=500$. This value of $p$ is
consistent with recent observations indicating that, the scalar spectral
index, $n_{S}$, which for a power-law background in the absence of minimal
length happens to be $1-2/p$, is greater than $0.95$ \cite{Spergel}. We also
assume that $\beta =100^{2}$, which corresponds to a minimal length $100$
times larger than Planck scale. This is a reasonable assumption in the
framework of scenarios of large extra dimensions \cite{Arkani-Hamed}. We see
that the standard tensor power spectrum is modulated by oscillations,
corresponding to a slow decrease in $H$ as the universe expands. Increasing $%
p$ (though still working with the rescaled variable $k$) does not change the
qualitative features of the power spectrum, it only results in a shift of
the $\log k$ axis to the left. Since $k$ appears in the mode equation as $%
k^{2/p}$ and $\tilde{k}_{crit}$ scales as $p^{p}$ this rescaling of
the axis can have a significant effect on the spectrum when we
compare power spectra for different $p$ with a common set of units
for $\tilde{k}$. As $p$ increases, the wavelength of the
oscillations increases \cite{Easther3}. Also, as $k$ increases the
frequency of the oscillation increases, though the amplitude
decreases. As expected, when $k\rightarrow \infty $, we recover the
standard field theory result. The left graph in Figure \ref{fig1}
illustrates the power spectrum for the modes that have a larger
amplitude than those seeding the structure formation in Hubble
patch. In the right graph we plot a window of $k$ where the
amplitudes are of the same order as the modes that are precursors to
structure formation, $10^{-5}\leq
P_{S}^{1/2}\leq 10^{-4}$ \cite{Easther3}, or equivalently with $p=500$, $%
10^{-7}\leq A_{T} \leq 10^{-6}$.

\begin{figure}[t]
\includegraphics[angle=0,
scale=0.80]{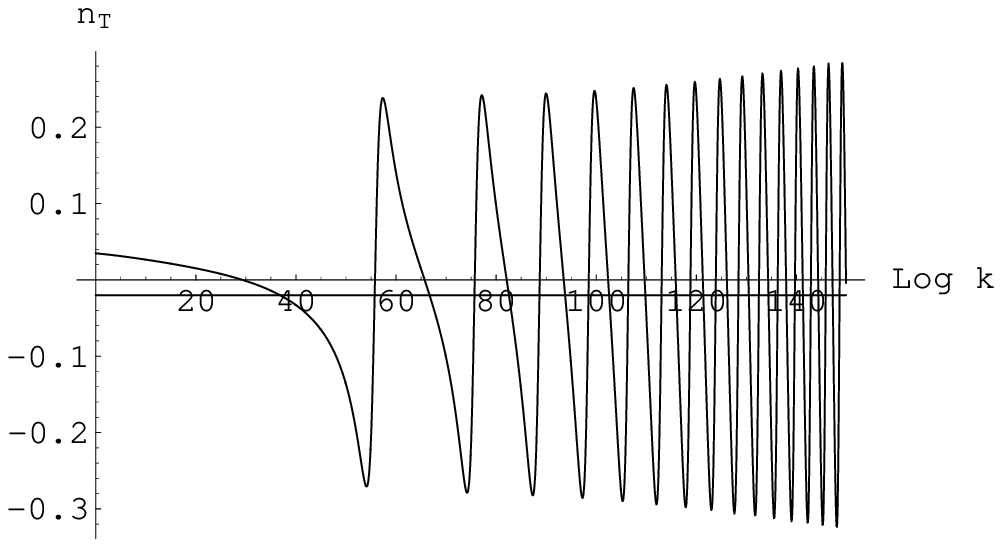}
\includegraphics[angle=0,
scale=0.80]{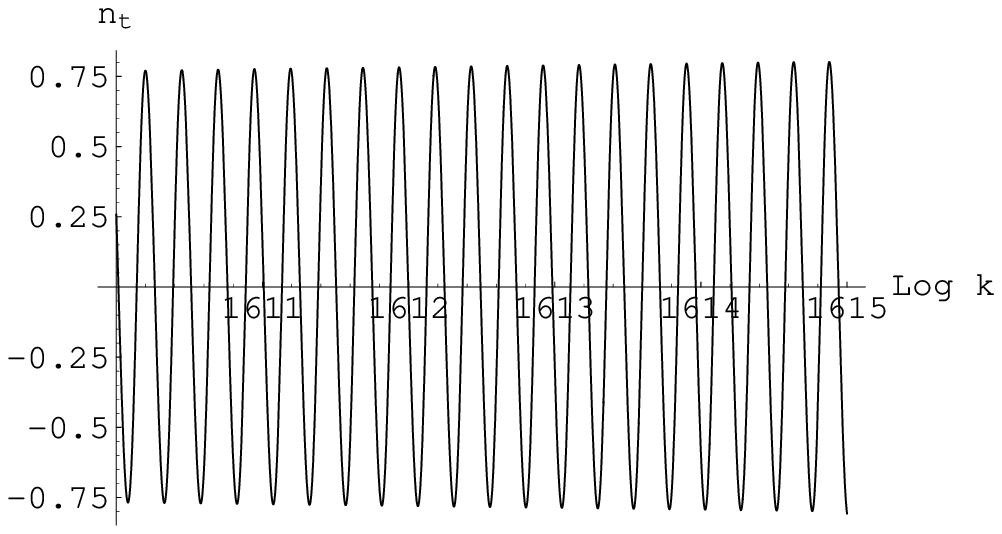}
\caption{$S_{T}^{(1)}$ is assumed to be the action of tensor fluctuations.
Left graph shows the dependence of $n_{T}$ on $\log k$ for wavelengths far
bigger than our horizon. $\protect\beta $ and $p$ are assumed to be $10^{4}$
and $500$, respectively. The horizontal line represents the result when
there is no minimum length. On the right we have graphed $n_{T}$ in the
observable range of $k$.}
\label{fig2}
\end{figure}

On the left in Figure 2 we plot the tensor spectral index for the range of
wavelengths that lie outside our horizon. The existence of minimal length
yields running from a blue to a red spectrum on such scales. This happens
despite the fact that $\epsilon $, the first slow roll parameter, does not
have a local minimum. This is a counterexample to generic result of \cite%
{Chung}, which claimes that if the spectral index is to run from a blue to a
red spectrum there must be a local minimum in the slope of the potential. On
the right in Figure \ref{fig2}, we graph $n_{T}$ in the observable range of $%
k$. While we see the expected oscillations about the standard value,
the large $k$ behavior is now more difficult to understand. The
increasing frequency of oscillations for the tensor power spectrum
results in a growth of both the amplitude and frequency of the
oscillations. Current measurements put a lower bound of $40$ on $p$
\cite{Spergel}. With such a weak lower bound, the frequency of the
oscillations is very small. As any measurement of the spectral index
is taken over a finite range of $k$, one would not be able to detect
such oscillations. Taking small intervals centered at successively
larger values of $k$ we would find that the average value of $n_{T}$
over the interval approaches the standard field theory result. To be
able to detect these oscillations one needs extremely precise
measurements. This oscillationary behavior of the spectral index is
quite distinct from another model of trans-Planckian physics based
on the non-commutativity of physical time and space coordinates
\cite{Bran-Ho}. For such such a model, it was shown that the
spectral index runs from $n>1$ on
large scales to $n<1$ , where transition happens on scales close to $%
H_{0}^{-1}$ \cite{Huang1, Huang2,Huang3,Maartens}.

Assuming that the action for scalar perturbations is described by $%
S_{S}^{(2)}$, the scalar modes satisfy eq.(\ref{tsr2-eq}). Exploiting the
dominant balance technique, we again extract the most singular terms in the
mode equation in the vicinity of the irregular singular point with the
approximate solution:
\begin{equation}
u_{k}(y)=C_{+}~F(k,y)(1+\epsilon _{1}(k,y))(1+\epsilon
_{2}(k,y))+C_{-}~F^{\ast }(k,y)(1+\epsilon _{1}^{\ast }(k,y))(1+\epsilon
_{2}^{\ast }(k,y)),  \label{sol-S2}
\end{equation}%
where
\begin{eqnarray}
F(k,y) &=&y^{3/4}H_{-3/4}(2\sqrt{B_{k}y})  \notag  \label{sol2} \\
\epsilon _{1}(k,y) &=&-\frac{ek^{2/p}}{6}(-qy)^{3/2}(2-3\log y) \\
\epsilon _{2}(k,y) &=&-\frac{qy^{2}}{48}\left(
48(1-q)+28iqe^{3/2}k^{3/p}+3ek^{2/p}(4+59q-14q\log y\right) .  \notag
\end{eqnarray}%
and $B_{k}$ is given by%
\begin{equation}
B_{k}=-\frac{q}{4}ek^{2/p}.  \label{Bk}
\end{equation}%
Again, we have a constraint on the integration constants $C_{+}$ and $C_{-}$
from the Wronskian condition:
\begin{equation}
|C_{+}|^{2}-|C_{-}|^{2}=-\eta _{\tilde{k}}\pi \sqrt{-q}e^{-3/2}.
\label{wronskian2}
\end{equation}%
In the rest of the analysis, we choose $C_{-}=0$, to have a
Bunch-Davies-like vacuum. However, we emphasize that this choice is not
unique and there is still a considerable amount of freedom in the choice of $%
C_{-}$. Specifically, inspired by our analysis in near-de-Sitter space \cite%
{tensor/scalar}, we conjecture that if
\begin{equation}
\lim_{\beta \rightarrow 0}\frac{C_{-}}{C_{+}}=0,  \label{lim}
\end{equation}%
we recover the standard result.

\begin{figure}[t]
\includegraphics[angle=0,
scale=0.80]{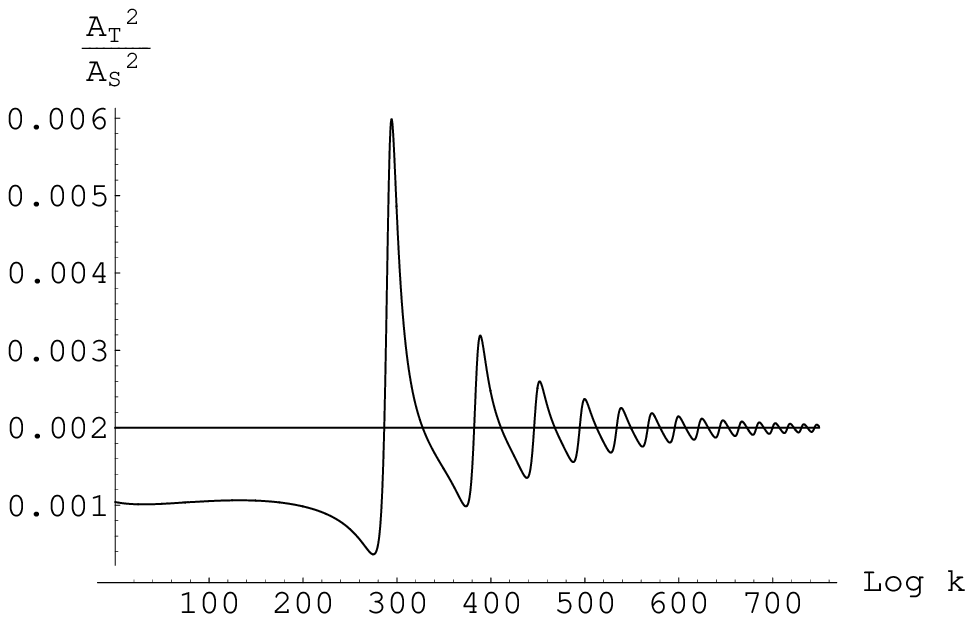}
\includegraphics[angle=0,
scale=0.80]{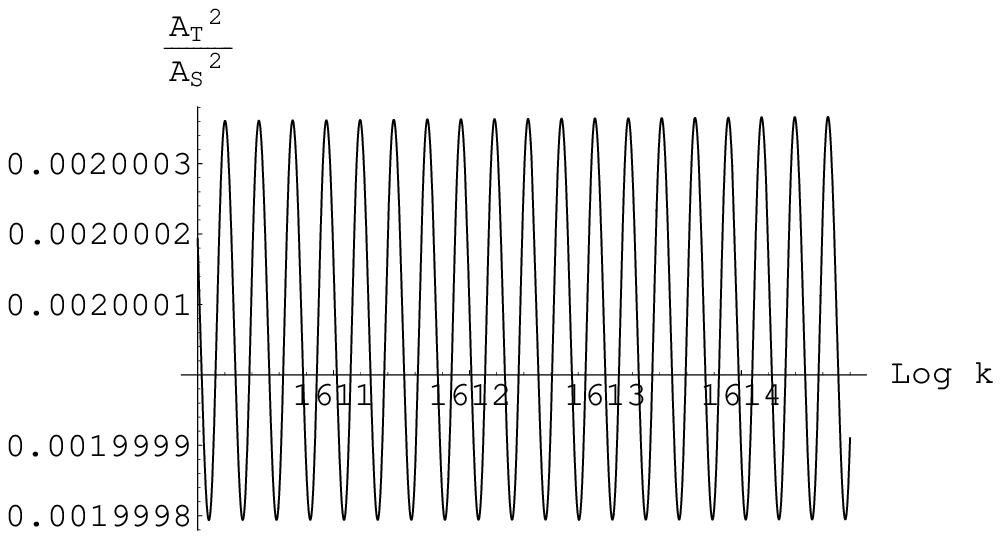}
\caption{We assume $(S_{T}^{(1)},S_{S}^{(2)})$ respectively describe tensor
and scalar fluctuations. The left figure shows $r$ for $p=500$ and $\protect%
\beta =100^{2}$ for wavelengths far bigger than our horizon. On the right we
plot the ratio of tensor to scalar fluctuations, in the observable window of
$k$.}
\label{fig3}
\end{figure}

This approximate solution is again used to set the initial conditions for a
numerical integration of the mode equation. The qualitative behavior of the
scalar power spectrum is found to be similar to that found for the tensor
modes. In Figure \ref{fig3} we have plotted the tensor/scalar ratio for $%
p=500$ and $\beta ^{1/2}=100$. The main effect of the different action for
the scalar perturbations in this case appears to be a slight ``compression''
of the oscillations to smaller $k$. This compression causes the
tensor/scalar ratio to oscillate in the observable window of $k$ about the
constant value we expect to find when there is no minimum length. This is
depicted on the right graph in Figure 3. Knowing this ratio is important if
one is to understand the contribution that each of these types of
perturbation makes to the anisotropy of the CMBR \cite{Salopek}. In Fig.\ref%
{fig3} we see from the left graph that the ratio stays constant at a value
less than the standard QFT result for small values of $k$ that correspond to
wavelengths outside our horizon. For increasing $k$ it attenuates until
reaches a minimum, after which it increases to a value much higher than the
standard result. Thereafter it starts oscillating about the standard QFT
predictions. The amplitude of the oscillations dies off as $k$ increases.
Notice also that $A_{T}^{2}/A_{S}^{2}$ is suppressed relative to $n_{T}$ by
an order of magnitude, implying in general a violation of the consistency
relation (\ref{consis}).

This behavior for the tensor/scalar ratio was anticipated from our earlier
calculations in near de-Sitter background \cite{tensor/scalar}, using $%
S_{T}^{(1)}$ for tensor and $S_{S}^{(2)}$ for scalar perturbations. In near
de-Sitter space for small values of $\sigma \equiv \sqrt{\beta }H$, the
ratio oscillates around the quantum field-theoretic prediction. For a fixed
value of minimal length, this corresponds to small values of the Hubble
parameter. As in a power-law background, short wavelength modes (large $k$)
experience a slower rate of expansion, and so we expect oscillationary
behavior in this region. Larger wavelengths are generated at the beginning
of inflation, when the Hubble parameter (and in turn $\sigma $) were larger.
For such wavelengths, this ratio is almost constant in a near de-Sitter
background. In a power-law background for such values of $k$ we see (Fig.2)
that the ratio is constant.

\subsection{$(S_{T}^{(2)},S_{S}^{(1)})$}

In this section, we assume that tensor and scalar perturbations satisfy eqs.(%
\ref{tsr2-eq}) and (\ref{scr1-eq}) respectively. In a power-law background, $%
z^{\prime \prime }/z=a^{\prime \prime }/a$ \thinspace\ \cite{Lidsey} and $%
z^{\prime }/z=a^{\prime }/a$ \cite{tensor/scalar} so the equation describing
scalar (tensor) perturbations is the same as the one describing tensor
(scalar) perturbations in section 4.1. From equation (\ref{r-slowroll-gen}),
one can deduce that $r/\epsilon $ now is just the inverse of $r/\epsilon $
from the last section.

In Figure \ref{fig4} we show the tensor spectral index derived from the
action $S_{T}^{(2)}$ overlaid on that found from $S_{T}^{(1)}$. Again, we
note that the removal of the third term in parentheses of (\ref%
{mode_eq_explicit}) causes a compression of the oscillations to smaller
values of $k$, but the magnitude of oscillations is still larger than those
of $A_{T}^{2}/A_{S}^{2}$ by an order of magnitude, indicating in general
that the consistency condition is still violated.

\subsection{$(S_{T}^{(1)},S_{S}^{(1)})$ and $(S_{T}^{(2)},S_{S}^{(2)})$}

For both of these cases the mode equations for tensor and scalar
fluctuations are identical. We therefore recover the standard field theory
result for the ratio $A_T^2/A_S^2$. The tensor spectral index has already
been presented in Figures \ref{fig1} and \ref{fig3}. Again the oscillations
about the standard result indicate there are violations of the consistency
condition.

\subsection{$\protect\beta$ dependence of fluctuations}

Up until now we have been working with a rescaled variable that allows us to
study the generic behavior for any $\beta$. Recall that the definition of
our variable $k$ involves $\beta$ dependence of the form $k \sim
\beta^{(p-1)/2} \tilde{k}$ and there is an overall factor of $\beta^{-1/2}$
in the normalization of the power spectrum. One may then qualitatively
compare our results for different values of $\beta$ by noting that, up to
normalization, changing $\beta$ just corresponds to a shifting of the $\log
k $ axis. For example, a value of $\beta=100^2$, causes the spectra of
Figures \ref{fig1}, \ref{fig2} and \ref{fig3} to shift to the left relative
to the $\beta=500^2$ results. For a given $\tilde{k}$ the net result is that
the size of the fluctuations about the standard field theory result are
suppressed.

\begin{figure}[t]
\includegraphics[angle=0,
scale=0.80]{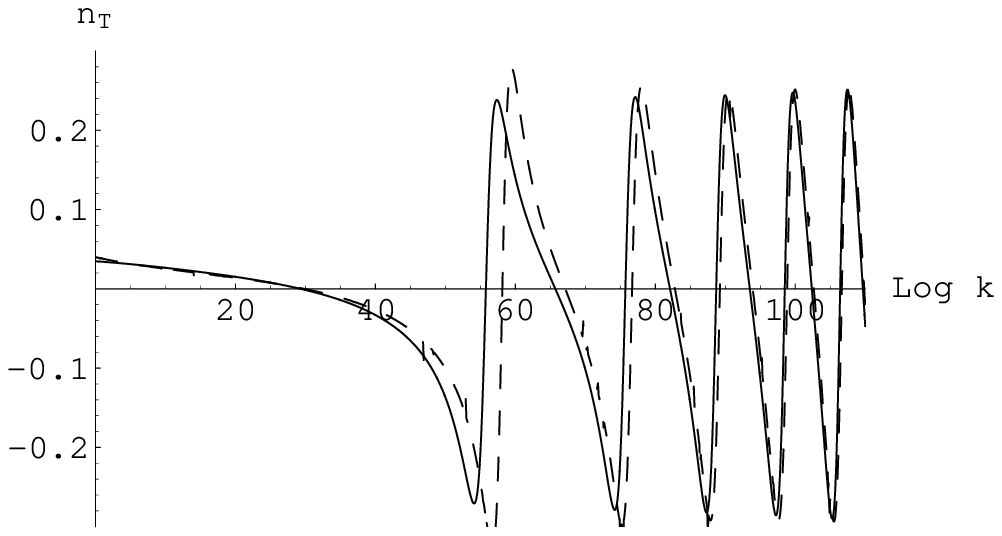}
\includegraphics[angle=0,
scale=0.80]{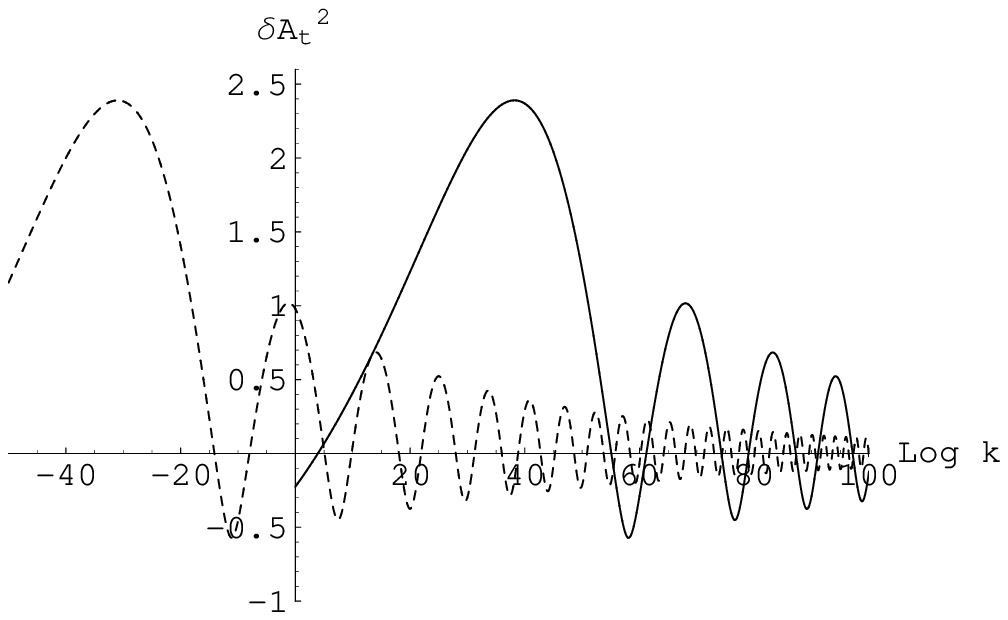}
\caption{In the figure on the left, we overlay the tensor index for the
power spectra obtained from $S_{T}^{(1)}$ (solid) and $S_{T}^{(2)}$
(dashed). On the right, we have graphed $\protect\delta A_{T}$,
modifications to the standard quantum field theory prediction due to the
presence of minimal length, for $\protect\beta =500^{2}$ (solid line) and $%
\protect\beta =100^{2}$ (dashed). Here, we have assumed that $S_{T}^{(2)}$
describes the action for tensor perturbations and $p=100$. }
\label{fig4}
\end{figure}

To be more exact, we may parameterize the tensor power spectrum as $A_T^2 =
A_{T,\mathrm{qft}}^2(1+\delta A_T(\beta,k))$, where $A_{T,\mathrm{qft}}$ is
the standard quantum field theory result for the tensor power spectrum. In
Figure \ref{fig4}, for action $S_{T}^{(2)}$, we plot $\delta A_T(\beta,k)$
for $\beta=500^2$ and $\beta=100^2$ written in units where $k=1 $
corresponds to $\tilde k = e^{p/2}/\tilde{k}_{crit}$ for $\beta=100^2$. We
find that the size of the oscillations and their wavelength both appear to
vary as $\beta^{1/2}$, the only dimensionful parameter in the problem.

\section{Conclusion}

In this article we investigated the consistency relation between tensor and
scalar fluctuations in the framework of power-law inflation with a cut-off
due to minimal length. Since the method of implementing the minimal length
hypothesis (\ref{commutation}) depends on the action one starts from, there
is a choice amongst an infinite number of actions that in the absence of
minimal length are otherwise equivalent. However there are only two
physically reasonable cases for both scalar and tensor perturbations: that
of minimality (add no boundary terms to the original action) and that of
simplicity (add terms such that the modified action most closely resembles
the action of a free massive scalar field in a Minkowski background). This
yields four distinct cases and we investigated each for a choice background
consistent with recent observations that constrain the magnitude of the
scalar spectral index.

Confining our attention to these cases, we found that Planck scale physics
can considerably modify the consistency condition (\ref{consis}) and can
lead to the running of spectral indices regardless of which action one
employs. Depending on the choice of action for tensor and scalar
perturbations, we may find that the tensor/scalar ratio oscillates (in the
observable window of $k$) about the constant value we expect to find in the
absence of minimal length. However the magnitude of the modifications
depends upon the choice of action. Constraining this choice -- both
observationally and theoretically -- remains a challenge for future studies.

\section{Acknowledgement}

We are thankful to A. Kempf, R. Easther and W. H. Kinney for helpful
discussions. This work was supported by the Natural Sciences \& Engineering
Research Council of Canada.


\begin{thebibliography}{99}
\bibitem{Brandenberger1} Robert H. Brandenberger, hep-ph/9910410.

\bibitem{Linde} A. D. Linde, \textit{Particle Physics and Inflationary
Cosmology}~(Harwood Academic, Chur, Switzerland).

\bibitem{Brandenberger2} J. Martin \& R. H. Brandenberger, Phys. Rev.
\textbf{D63}, 123501 (2001), hep-th/0005209; R. H. Brandenberger \& J.
Martin, Mod. Phys. Lett. \textbf{A16}, 999 (2001), astro-ph/0005432; J.
Martin, R. H. Brandenberger, Phys. Rev. \textbf{D65}, 103514

\bibitem{Greene} B. Greene, K. Schalm, G. Shiu, J. P. van der Schaar,
astro-ph/0503458

\bibitem{Unruh} W. Unruh, Phys. Rev. \textbf{D51}, 2827 (1995)

\bibitem{Corely} S. Corely \& T. Jacobson, Phys. Rev. \textbf{D54}, 1568
(1996), hep-th/9601073; S. Corely, Phys. Rev \textbf{D57}, 6280 (1998),
hep-th/9710075

\bibitem{Riotto} D. J. H. Chung, E. W. Kolb \& A. Riotto, Phys.\ Rev.\
\textbf{D65}, 083516 (2002) hep-ph/0008126

\bibitem{Volovik} G. E. Volovik, Phys.\ Rept.\ \textbf{351}, 195 (2001),
gr-qc/0005091; G. E. Volovik, Pisma Zh.\ Eksp.\ Teor.\ Fiz.\ \textbf{73},
182 (2001) [JETP Lett.\ \textbf{73}, 162 (2001)], hep-th/0101286

\bibitem{Brandenberger3} R. H. Brandenberger \& J. Martin, Int.\ J.\ Mod.\
Phys.\ \textbf{A17}, 3663 (2002), hep-th/0202142

\bibitem{Starobinsky} A. A. Starobinsky, Pisma Zh.\ Eksp.\ Teor.\ Fiz.\
\textbf{73}, 415 (2001) [JETP Lett.\ \textbf{73}, 371 (2001)],
astro-ph/0104043

\bibitem{Tanaka} T. Tanaka, astro-ph/0012431

\bibitem{Danielsson} U. H. Danielsson, Phys. Rev. \textbf{D66}, 023511 (2002)

\bibitem{alberghi} G. L. Alberghi, R. Casadio, A. Tronconi, Phys. Lett.
\textbf{B579}, 1 (2004), gr-qc/0303035; R. Easther, B. R. Greene, W. H.
Kinney and G. Shiu, Phys. Rev. \textbf{D66}, 023518 (2002), hep-th/0204129

\bibitem{Kempf1} A. Kempf, Phys. Rev. \textbf{D63}, 083514 (2001),
astro-ph/0009209

\bibitem{Kempf2} A. Kempf, hep-th/9810215

\bibitem{Gross} D.~J.~Gross and P.~F.~Mende, Nucl.\ Phys. \textbf{B303}, 407
(1988). ; D. Amati, M. Ciafaloni, G. Veneziano, Phys. Lett. \textbf{B216} 41
(1989)

\bibitem{Easther2} R. Easther, B. R. Greene, W. H. Kinney \& G. Shiu, Phys.
Rev. \textbf{D64}, 103502 (2001), hep-th/0104102

\bibitem{Easther3} R. Easther, B. R. Greene, W. H. Kinney \& G. Shiu, Phys.
Rev. \textbf{D67} 063508 (2003)

\bibitem{Ambiguity} A. Ashoorioon, A. Kempf, R. B. Mann, Phys. Rev. \textbf{%
D71}, 023503 (2005), astro-ph/0410139

\bibitem{tensor/scalar} A. Ashoorioon, R. B. Mann, Nucl. Phys. \textbf{B716}%
, 261 (2005),  gr-qc/0411056

\bibitem{cosmag} Amjad Ashoorioon, Robert B. Mann, Phys. Rev. \textbf{D71},
103509 (2005), gr-qc/0410053

\bibitem{star-lyth} A. A. Starobinsky, Sov. Astr. Lett., \textbf{11}, 133
(1985); E. D. Stewart \& D. H. Lyth, Phys. Lett. \textbf{B302}, 171 (1993)

\bibitem{Kinney} L. Hui \& W. H. Kinney, Phys. Rev. \textbf{D65} 103507
(2002), astro-ph/0109107

\bibitem{Lidsey} J. E. Lidsey, A. R. Liddle, E. W. Kolb, E. Copeland, T.
Barreiro \& M. Abney, Rev. Mod. Phys. \textbf{69}, 373, (1997)

\bibitem{Corless} R. Corless, G. Gonnet, D. Hare, D. Jeffrey, and D. Knuth,
Adv. Comput. Math. \textbf{5}, 329 (1996).

\bibitem{Mukhanov} Sov.\ Phys.\ JETP \textbf{67}, 1297 (1988) [Zh.\ Eksp.\
Teor.\ Fiz.\ \textbf{94N7}, 1 (1988)]; V. F. Mukhanov, H. A. Feldman, R. H.
Brandenberger, Phys. Rep. \textbf{215}, 203 (1992)

\bibitem{Salopek} D. S. Salopek, Phys. Rev. Lett. \textbf{69} 3602 (1992).

\bibitem{Liddle} A. R. Liddle, D. H. Lyth, Phys. Lett. \textbf{B291} 391
(1992).

\bibitem{Bender} Carl M. Bender and Steven A. Orszag, ''Advanced
Mathematical Methods for Scientists and Engineers'', McGrawHill, Inc., 1978.

\bibitem{Arkani-Hamed} N. Arkani-Hamed, S. Dimopoulos \& G. Dvali, Phys.
Lett. \textbf{B429} 263 (1998); I. Antoniadis, N. Arkani-Hamed, S.
Dimopoulos \& G. Dvali, \textit{ibid} \textbf{436}, 257 (1998).

\bibitem{Spergel} D. N. Spergel, \textit{et. al.}, Astrophys. J. Suppl. 148
(2003) 175, astro-ph/0302209

\bibitem{Chung} D. J. H. Chung, G. Shiu \& M. Trodden, Phys. Rev. \textbf{D68%
} 063501, 2003, astro-ph/0305193

\bibitem{Bran-Ho} R.~Brandenberger and P.~M.~Ho, Phys. Rev. D \textbf{66},
023517 (2002), hep-th/0203119

\bibitem{Huang1} Q.~G.~Huang and M.~Li, JHEP \textbf{0306}, 014 (2003),
hep-th/0304203

\bibitem{Huang2} Q.~G.~Huang and M.~Li, JCAP \textbf{0311}, 001 (2003),
astro-ph/0308458

\bibitem{Huang3} Q.~G.~Huang and M.~Li, Nucl. Phys. \textbf{B713}, 219
(2005) astro-ph/0311378

\bibitem{Maartens} S.~Tsujikawa, R.~Maartens and R.~Brandenberger, Phys.
Lett. \textbf{B574}, 141 (2003), astro-ph/0308169
\end{thebibliography}
\end{document}